\documentstyle [aps,tighten,preprint] {revtex}
\def\be{\begin{equation}}
\def\ee{\end{equation}}

\newcounter{tempfigc}
\newcommand{\fcaption}[1]{
	\addtocounter{tempfigc}{1}
{\noindent\parbox{18.5 cm} {\small Fig.~\thetempfigc. #1} }}

\newcounter{temptabc}
\newcommand{\tcaption}[1]{
	\addtocounter{temptabc}{1}
{\noindent\parbox{18.5 cm} {\small Tab.~\thetemptabc. #1} }}

\begin{document}
\begin{titlepage}
\title{ BARYONS AS SOLITONS \\
      IN EFFECTIVE CHIRAL FIELD THEORIES$^1$}
\author {G. HOLZWARTH$^2$ \\
	{\it Fachbereich Physik, Universitt-GH-Siegen} \\
	{\it  57068 Siegen, Germany }\\}
\maketitle
\begin{abstract}
This lecture comprises some recent developments concerning the
description of baryons as topological
solitons in effective chiral meson theories.\\
In the first part one-loop corrections to the classical
tree approximation are discussed. This involves renormalization of
low-energy coupling constants and evaluation of the finite
next-to-leading-order terms in the $1/N_c$ expansion.
In contrast to the corresponding procedure in the meson sector
the magnitude of the chiral gradients involved in the soliton profile
requires that counter terms and finite loop contributions be
calculated to all chiral orders. Recent results for various
nucleon observables are presented. They show that the $1/N_c$ expansion
essentially works as expected.\\
In the second part electro-magnetic nucleon form factors (FFs) with
relativistic corrections are evaluated in a chiral soliton model
including vector mesons. The magnetic FF $G_M^p$ is shown to
agree well with new SLAC data for spacelike $Q^2$ up to 30
(GeV/c)$^2$ if superconvergence is enforced. The electric FF
$G_E^p$ is dominated by a zero in the few (GeV/c)$^2$ region
due to a low-lying zero in the non-relativistic electric FF in tree
approximation. \\
The third part describes how to extract the strong $\pi NN$ form factor
from chiral soliton models, taking due care of the local metric created
by the presence of the soliton. When used in a one-boson-exchange model
for the nucleon-nucleon (NN) interaction, deuteron properties and
phase parameters of NN scattering are reproduced
as well as in conventional NN models that apply a hard monopole
form factor at the $\pi NN$ vertex.
\end{abstract}
\vspace{2cm}
\begin{flushleft}
{\baselineskip = 14pt
hep-ph/9511249\\
November, 1995\\}
\end{flushleft}
\vfill
\footnotetext[1]{
Lecture presented at the International School of Nuclear Physics,\\
 17th Course:"Quarks in Hadrons and Nuclei"\\
 at the Ettore Majorana Centre for Scientific Culture, Erice, Italy,
September 19th-27th, 1995.}
\footnotetext[2]{e-mail: holzwarth@hrz.uni-siegen.d400.de}
\end{titlepage}
\section{Introduction}
Topological soliton models for structure and dynamics of baryons
are based on effective nonlinear lagrangians for selected
mesonic degrees of freedom. These usually comprise the set
$U(\mbox{\boldmath $x$},t)$ of pseudoscalar Goldstone
bosons of spontaneously broken chiral symmetry,
but also the light vector meson octet, and axial vector mesons have been
included.

Chiral Perturbation Theory (ChPT)~\cite{Wei79,GL} provides
a systematic way to establish order by order the effective
lagrangian ${\cal L}(U)$ for the pseudoscalar chiral Goldstone bosons.
To the extent that the relevant coupling constants can be
fixed with sufficient accuracy for a consistent description of
low-energy mesonic processes, the resulting part of the
effective lagrangian ideally should coincide with the chiral lagrangian
underlying the soliton model for baryons.
This must be so because far away from the center of a soliton the
fluctuations of the chiral field describe mesons which should interact
in the same way as the mesons in the vacuum sector.

It is always the first step in the actual use of a chiral effective
lagrangian to construct a static solution $U_0(\mbox{\boldmath $x$})$
which minimizes the classical effective action .
Depending on boundary conditions, chiral
lagrangians allow for topologically distinct sets of classical solutions,
characterized by suitable integer winding numbers $n$.
Following Skyrme~\cite{Sk61} and Witten~\cite{Wi83},
the identification of the
topological index $n$ with baryon number $B$, establishes a close link
between the concepts of ChPT and soliton models for baryons.

Although it is quite remarkable that the presently known low-energy
coupling constants of ChPT support stable static soliton solutions,
there still are crucial differences in the practical application
of both concepts:

In ChPT the scale of external momenta involved in a physical process
limits the chiral order $i_{\max}$ to which the effective lagrangian
${\cal L}$  has to be considered:
\be
{\cal L}(U) = \int \sum_{i=2,4,6,..}^{i=i_{\max}} \;
\;c_i \; \left(L_\mu \right)^i \; d^3x
\ee
where $L_\mu$ denotes the chiral gradients
\be
L_\mu = U^\dagger \partial_\mu U .
\ee

For a soliton solution in the $B$=1 sector the gradients of the
soliton profile are of the order of $(\pi/d)$ where $d$ is the
typical size of the soliton. For $d\sim 1$ fm we have
$\pi/d\sim$ 700 MeV, i.e. the gradients typically are
of the order of the $\rho$-mass $m_\rho$.
This excludes a justification based on the power of chiral
gradients, for truncating the effective lagrangian to some low chiral
order. The limitation to low powers of $L_\mu$ in ${\cal L}$ which is
necessary for practical reasons, then implies the {\it assumption} that the
renormalized coupling constants $c_i$ for $i > i_{max}$ are so small
that effects of the omitted higher-order terms may be safely neglected.

Numerous investigations of the (chiral-order-four) Skyrme model
and higher-order extensions have shown, that chiral profile and
baryonic properties in tree approximation do not differ drastically, if
only the size of the soliton is of comparable magnitude. In fact, to
some extent this holds even for chiral lagrangians with vector mesons
included, which effectively sum up terms to arbitrarily high chiral
orders~\cite{MKW87,JJMPS88,MKWS89,SWHH89}.

Going beyond the classical tree approximation requires evaluation of the
chiral field fluctuations $\pi(\mbox{\boldmath $x$},t)$ around
the static soliton $U_0(\mbox{\boldmath $x$})$
\be
U(\mbox{\boldmath $x$},t)=
U_0(\mbox{\boldmath $x$}) \exp [ i \pi(\mbox{\boldmath $x$},t)/f_\pi] \;.
\ee
One-loop corrections only involve the interaction $V$
of the fluctuations with the soliton
background field; higher loops involve also their mutual interactions.
It is essential that the hamiltonian $h_S^2=h_0^2+V$ which
underlies the dynamics of these fluctuations
be consistent with the equation of motion which determines the
static soliton itself. (Otherwise zero modes will no longer be zero modes,
terms linear in the fluctuational fields will not vanish, and the
meson-soliton interaction $V$ will depend on the choice of the
parametrization chosen for the fluctuational field. )
Therefore in $h_S^2$ all terms have to be kept which arise from the
full effective lagrangian used in tree approximation to evaluate the
classical soliton.

Renormalization of loop corrections~\cite{Mou,Ho94,HoWa95}
then requires counter terms which already in one-loop
approximation renormalize the parameters
of {\it all} higher chiral orders in the effective lagrangian ${\cal L}$.
Still, as stated for the meson sector in Weinberg's power-counting
rule~\cite{Wei79}, the renormalized coupling constants for the terms of some
given chiral order $i$ pick up loop contributions only from those
terms in $\cal {L}$ with chiral order $i-2$ and less. This allows for
the important fact that renormalization in the meson sector
simultaneously renormalizes also the soliton sector.
However, the resulting finite
loop-corrections to physical quantities involve {\it all} powers
of the terms kept in the lagrangian and cannot be truncated because
the gradients involved are not small. This is an important difference
to the procedures applied in the meson sector and it rules out early
attempts~\cite{ZWM86} to calculate loop corrections in the soliton sector by
counting chiral orders in the same way as in the meson sector.

So, to restate the essentials: i) The renormalization of
loops in the soliton sector leads to the {\it same} renormalized
coupling constants as in the meson sector.
ii) The finite loop
contributions to the observables of interest contain {\it all} powers of
the terms which are kept in the effective lagrangian, i.e. they
comprise {\it all} chiral orders.
iii) A truncation of the effective lagrangian
can {\it not} be justified on grounds of increasing chiral order,
but it implies the {\it assumption}
that the renormalized coupling constants for the
omitted terms are very small. (Such a statement of course is scale
dependent).

{}From the foregoing it is quite evident that using the established
parts of the effective chiral lagrangian in the soliton sector
lacks the rigor of the systematic ChPT scheme, while going beyond the
established parts turns this fundamental tool into a mere
QCD-inspired {\it model} for baryon structure and dynamics.
While in ChPT loop corrections are in tight correspondence
with higher orders in the chiral expansion,
in the soliton sector tree, one-loop, two-loop approximation etc.
merely represent the first terms in the $1/N_c$ expansion, with each
term comprising {\it all} chiral orders, independent of the maximal
chiral order $i_{max}$ kept in the effective action.
Thus, evaluation of one-loop corrections in the soliton sector
provides information not only about the magnitude of these $1/N_c$-
corrections, but an eventually remaining scale dependence of physical
observables indicates how seriously the lack of rigor may affect
the results.

\section{An example: The nucleon mass}
As an example let us consider the nucleon mass in more detail.
In an effective field theory
characterized by a classical (Euclidean) action $S^E$
the mass of the lowest quantum eigenstate is given by~\cite{Raja}
\be
M=-\lim_{\tau \to \infty}\frac{1}{\tau} S^E
\ee
where $\tau$ limits the (imaginary-)time integral in the
action $S^E$. Tree and one-loop approximation are the
leading and next-to-leading-order terms in a $1/N_c$ expansion
of the action
\be
S^E = S_0^E -\frac{1}{2}\, \mbox{ Tr  ln}\, D^E \;+\cdot \cdot\cdot
\ee
where $S_0^E$ is the action taken at the static classical solution
and $D^E$ is the operator which determines the harmonic fluctuations
around the classical minimum.
For chiral lagrangians the classical static configuration is the
hedgehog soliton while the fluctuations asymptotically correspond
to the pseudoscalar mesons.
The scale in the meson-soliton interaction is set by gradients of the
soliton profile which as we discussed above are of the order of
of the $\rho$-mass $m_\rho$. Generally, therefore,
it is not sufficient to evaluate the fluctuations in Born approximation.
This is consistent with the fact that evaluation of $\pi N$-scattering
phase shifts involves the pion-soliton interaction $V$ to all
orders, irrespective of the maximal power $i_{\max}$ of chiral derivatives
which occur in the effective lagrangian. Similarly, it is not possible
to obtain the zero modes in Born approximation; however, they provide
the most important part of the quantum corrections to the baryon mass.

After performing the time integral in $S^E$ the baryon mass then is given
as
\be
M=M_S+E_{cas}
\ee
with ${\cal O}(N_c)$ classical soliton mass $M_S$ and
${\cal O}(1)$ Casimir energy
\be
E_{cas}=\frac{1}{2}(tr\;h_S - tr\;h_\infty)
\ee
The hamiltonian $h_S$ governs the dynamics of the soliton fluctuations.
Asymptotically, far away from the soliton, $h_S^2$ turns into the
kinetic energy operator $h_\infty^2=\mbox{\boldmath $p$}^2 +m_{\pi}^2 $
for free pions. In the Casimir energy the infinite ${\cal O}(1)$ vacuum
energy contribution is subtracted.

The Casimir energy is a formal object which requires renormalization.
In order to determine the relevant counter terms
(and {\it only} for that purpose) we separate $h_S^2$ into kinetic
part $h_0^2$ and chiral covariant momentum-independent interaction part $V$
\be
h_S^2=h_0^2+V
\ee
and consider the formal Taylor expansion of the square root of
$h_S^2$ in (7) in powers of the interaction $V$
\be
tr\;h_S = tr \; h_0 \sum_{k=0}^\infty \lambda_k
\left(\frac{V}{h_0^2}\right)^k,
\;\;\;\; (\lambda_0=1,\;\;\lambda_1=1/2,\;\;\lambda_2=-1/8, \;...)
\ee

The ultraviolet divergencies reside (for 3+1 dimensions) in the first
three terms $k=0,1,2$, while terms with $k>2$ are finite.
The series may be visualized diagrammatically as a sum of
one-pion-loops with $k$ insertions of the pion-soliton vertex $V$:

\setlength{\unitlength}{1mm}
\begin{picture}(150,25)
\put(50,15){\circle{15}}
\put(45,0){\makebox(10,3){k=0}}
\put(61,13){\makebox(3,3){+}}
\put(75,15){\circle{15}}
\put(70,0){\makebox(10,3){k=1}}
\put(86,13){\makebox(3,3){+}}
\put(100,15){\circle{15}}
\put(95,0){\makebox(10,3){k=2}}
\put(111,13){\makebox(3,3){+}}
\put(125,15){\circle{15}}
\put(120,0){\makebox(10,3){k=3}}
\put(136,13){\makebox(3,3){+}}
\put(141,13){\makebox(10,3){. . .}}
\put(67.5,15){\circle*{2}}
\put(100,22.5){\circle*{2}}
\put(100,7.5){\circle*{2}}
\put(117.5,15){\circle*{2}}
\put(129.5,20){\circle*{2}}
\put(129.5,10){\circle*{2}}

\end{picture}
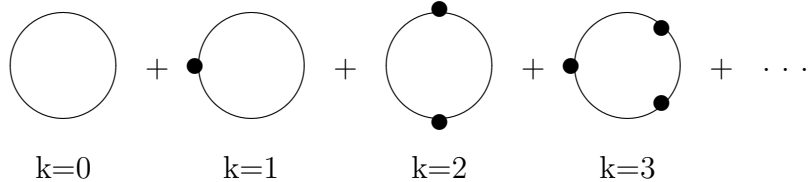
\fcaption{~~~~~~~~~~~~~~~~~ Graphical representation of
the one-loop multi-vertex series (9).}

There is no small expansion parameter in this series and as we have
stressed before, it must be summed to all orders.
Formally, the renormalization procedure consists of the following:
Subtract from $E_{cas}$ counterterms which
remove the infinities in the series (9) and add them again:
\be
E_{cas}=\frac{1}{2} tr(h_S-h_0-\frac{1}{2 h_0}V
      +\frac{1}{8 h_0^3}V^2)
+\frac{1}{2}(\alpha_0 a_0+\alpha_1 a_1+\alpha_2 a_2)
\ee
with the divergent integrals
\be
\alpha_k =  \lambda_k/(2\pi)^3 \int \!d^3p\,(\mbox{\boldmath $p$}^2 +
m_\pi^2)^{1/2-k},
\ee
which are conveniently handled in dimensional regularization with
renormalization scale $\mu$. In the ($\bar{MS}$) subtraction scheme
the infinite parts of the integrals $\alpha_k$ are absorbed into
scale-dependent renormalized coupling constants $c_i(\mu)$
in the (sufficiently general) effective lagrangian
(this is always possible because the spatial integrals $a_0,a_1,a_2$
are chirally covariant). Remaining scale-dependent finite
parts of the $\alpha_k$ integrals, together with
the scale-independent finite part of $E_{cas}$ in (10) then constitute
the scale-dependent finite one-loop correction $E_{cas}(\mu)$ to
the classical soliton mass $M_S(\mu)$ (which is implicitly
scale dependent via the renormalized coupling constants $c_i(\mu)$).

Exact evaluation of $tr\,h_S$ requires knowledge of the
exact scattering eigenstates for $h_S^2$. They have been calculated
previously~\cite{WE84} for Skyrme- and similar models, so there is
no principal difficulty in
obtaining precise numerical results.

\section{Some results and remarks about the $1/N_c$ -expansion}
At present results for one-loop corrections have only been obtained
in the framework of chiral SU(2)$\times$SU(2), i.e. for purely pionic
dynamics. In this case the chiral-order-four part of ${\cal L}$ consists
(apart from small terms proportional to the pion mass) of only two
terms, the antisymmetric Skyrme term
\be
{\cal L}_A^{(4)}=\frac{1}{32e^2}\int tr[L_\mu,L_\nu]^2 d^3x
\ee
and the symmetric term
\be
{\cal L}_S^{(4)}=\frac{\gamma}{8 e^2}\int (tr L_\mu L_\mu)^2 d^3x
\ee
in addition to the usual (chiral-order-two) part
\be
{\cal L}^{(2)}=\frac{f_\pi^2}{4}\int \left(-trL_\mu L^\mu+m_\pi^2
tr(U+U^\dagger-2) \right)d^3x.
\ee

In ChPT, with the conventional choice for the renormalization scale
$\mu=m_\rho$, the magnitude of the Skyrme parameter $e$ has been determined
to lie in the range 5.8 $< e <$ 7.2, while $\gamma$ is compatible with zero.
As we have remarked already, this order-four lagrangian stabilizes static
soliton solutions. However, for a reasonable size of the resulting baryon,
values of the Skyrme parameter around $e \approx 4$ are necessary. This is
a clear indication that for a reliable description of baryons in the soliton
sector more terms in the chiral expansion of ${\cal L}$ are required.
There are, however, so many possible types of terms all with unknown
coupling constants, that presently one reasonable strategy seems
to be restriction to ${\cal L}^{(4)}$ with a choice for $e$ which
differs from its value determined in ChPT, but effectively represents
the influence of the higher-chiral-order terms on the soliton solutions.
A posteriori, this compromise may be justifiable if it turns out that
the one-loop corrections lead to significant improvement for many
baryonic observables and, as we have discussed earlier, that the results
show only a mild dependence on the renormalization scale $\mu$.

Fig.2 shows results obtained for the baryon mass with $e$=4.25 and
$\gamma$=0 at $\mu=m_\rho$. Through the scale dependence of the
renormalized coupling constants the classical soliton mass $M_S(\mu)$
varies between 1400 and 1800 MeV in the scale range of
450 MeV $< \mu <$ 1200 MeV. Remarkably, the one-loop correction
compensates this dependence to such a degree that in the scale range
from the $\eta$-mass $m_\eta$ up to 1 GeV the resulting baryon mass
is constant near 946 MeV with very good accuracy. Below $\mu \approx$ 500
MeV the value of $\gamma(\mu)$ approaches the limit where the symmetric term
(13) destabilizes the classical soliton solution.

Similarly, numerical results can be obtained for other observables by
recalculating tree and one-loop contributions to the mass in presence
of suitably chosen external fields. Evaluating derivatives of the baryon
mass with respect to the strengths of the imposed external fields provides
magnitude and scale dependence for any desired observable in leading and
next-to-leading $1/N_c$-order. The same parameter choice as in fig.2 (i.e.
$e$=4.25, $\gamma$=0 for $\mu$=$m_\rho$) leads to the following results:
(the first number for a given observable is the classical
${\cal O}(N_c)$ result; the second number is the ${\cal O}(1)$
one-loop correction; standard experimental values are given in brackets)

{}~~~~~~~baryon mass: $  M=1628-682=946\;\;\;(939)$ [MeV]

{}~~~~~~~$\pi$-N sigma term: $\sigma=54-21=33 \;\;\;(45)$ [MeV]

{}~~~~~~~isovector magn. moment: $\mu_V=1.62+0.62=2.24 \;\;\;(2.35)$

{}~~~~~~~isovector magn. square radius:
$<r^2_M>_V=0.86-0.14=0.72\;\;\;(0.73)$ [fm$^2$]

{}~~~~~~~proton polarizability:
$\alpha_p=17.8-8.0=9.8\;\;\;(9.5)$ [$10^{-3}$ fm$^3$].

These numbers, quoted from Meier and Walliser~\cite{MW95}, allow to conclude
that, first of all, the quantum corrections to the classical results
are not small. In fact, although we expect them as
$1/N_c$ corrections, their magnitude in general exceeds the
30$\%$ estimate (depending on the chosen scale). Still, cum grano salis,
they seem to confirm the applicability of the $1/N_c$-expansion, and
the fact that they do improve the general agreement with experimental
baryonic properties, supports our choice of an effective chiral-order-four
lagrangian.

The corresponding consideration for the axial coupling constant $g_A$
shows, however, that for certain observables the $1/N_c$-expansion
can be more problematic than the numbers given above may suggest.
For the same parameter choice the classical ${\cal O}(N_c)$
result for $g_A$ at $\mu=m_\rho$ is $g_A$=0.90. At first sight, this
appears quite in line with our expectation of a 30$\%$ ${\cal O}(1)$
correction to get close to the experimental value of 1.25. However,
the constraint imposed on $g_A$ by the Adler-Weisberger relation
\be
g_A^2=1+\delta,  \;\;\;\;\;\; \delta>0,
\ee
requires that $g_A^2$ must contain ${\cal O}(N_c^0)$ contributions with a
numerical magnitude of about 1, while $g_A^2$ which is of
${\cal O}(N_c^2)$ has a numerical value of about 1.5.
This shows that in order to reproduce the experimentally observed number for
$g_A$ in a $1/N_c$ expansion it must be calculated at least to
${\cal O}(1/N_c)$ which is {\it two}
orders down from the leading ${\cal O}(N_c)$,
and the resulting corrections to $g_A^2$ (or to $\delta$) will be of
the same order of magnitude
as the leading term. So, if the tree approximation would happen to be
close to the experimental result, one could even expect the one-loop
corrections to take it away from this fortunate value to make room
for the necessarily large two-loop contributions.
And this is what happens with the above choice of parameters:

{}~~~~~~~~~~axial coupling constant: $g_A=0.90-0.25=0.65 \;\;\;(1.25)$

The origin of this difficulty is located in the current algebra, which
requires the commutator of two axial currents (which both are of
${\cal O}(N_c)$) to be equal to a vector current
(which is of ${\cal O}(1)$), and it is the nucleon matrix element of this
vector current which is responsible for the 1 on the right-hand side
of the Adler-Weisberger relation~\cite{RiKi}. So, perhaps we should not be too
disappointed that the long-standing trouble with $g_A$ in soliton models
still cannot be resolved at the present level.

\section{Relativistic form factors in soliton models}

A decisive advantage of the soliton concept as compared to all other models
where pointlike fermion fields are coupled to mesons or gauge fields
is the fact that already in leading classical approximation
the spatial structure of the baryon as an extended object
is obtained from the underlying effective action. Therefore all types of
form factors can readily be extracted from the model itself
and precise measurements of their $Q^2$-dependence present
a severe test for the resulting spatial profiles.
Specifically, for  electro-magnetic form factors (e.m.FF)
new SLAC data~\cite{Si93,SLAC94} pose a challenge for
the relativistically corrected FFs of chiral soliton models.
And in the few (GeV/c)$^2$ region we expect a wealth of precise data
from the new generation of electron accelerators.
In this region of momentum transfer it is important to compare data
with relativistically corrected FFs.

The implementation of relativistic corrections
is fairly easy for solitonic nucleons
due to the Lorentz covariance of the underlying field equations.
 The corrections reflect the Lorentz
boost from the soliton rest frame to the Breit frame, in which the
soliton moves with velocity $v$ which satisfies
\be
\gamma^2=(1-v^2)^{-1} = 1 + \frac{Q^2}{(2M)^2}
\ee
for momentum transfer $Q^2$ ($Q^2>0$ in the spacelike region) and
soliton mass $M$.
The classical result for the magnetic FF is~\cite{Ji91}
\be
G_M (Q^2) = \gamma^{-2}\; G^{nr}_M ( \gamma^{-2}\;Q^2) \; ,
\ee
where $G^{nr}$ is the nonrelativistic FF evaluated in the
soliton restframe.
The electric FF $G_E$
does not contain the factor $\gamma^{-2}$ on the right-hand side~\cite{Ji91}:
\be
G_E (Q^2) = G^{nr}_E ( \gamma^{-2}\; Q^2)
\ee
 (this is in contrast to bag models~\cite{boobag} where the wave functions of
the spectator quarks supply the factor $\gamma^{-2}$ also for $G_E$.)

According to the derivation of (17,18) within tree approximation of
the soliton model the kinematical mass $M$ in (16) is the
classical soliton mass $M_S$, although ideally, of course, $M$
should coincide with the physical nucleon mass $M_N$.
{}From (17,18) the asymptotic limit of $G(Q^2)$ for $Q^2 \rightarrow
\infty$ is determined by $G^{nr}(4M^2)$.
For commonly used chiral lagrangians the first zeros of the
nonrelativistic FFs occur at masses $M_0$
\be
G^{nr}(4 M^2_{0}) = 0
\ee
which are rather close to the nucleon mass, with $M_0<M_N$ for
$G_E^{nr}$ and $M_0>M_N$ for $G_M^{nr}$. This implies that the
asymptotic behaviour of $Q^4G(Q^2)$ is very sensitive to the precise
value of $M$ used in (16-18):
\be
\lim_{Q^2 \rightarrow \infty} Q^4G(Q^2) = \pm \infty
  \mbox{~~~~for~~~~}  M \begin{array}{c} <\vspace{-3mm}\\>\end{array} M_0.
\ee
The actual values of $M_0$ for which $G^{nr}(4M_0^2)$ vanishes, depend
on the choice of the parameters in the effective lagrangian;
furthermore both, $M_S$ and $G^{nr}$ are subject to quantum
corrections. It is therefore unrealistic to expect reliable predictions
from the model itself for the high-$Q^2$ behaviour of $Q^4G(Q^2)$.

However, this ambiguity in  the high-$Q^2$ behaviour of $Q^4G(Q^2)$
can be used to {\it {impose}} superconvergence
$( Q^2 G_M (Q^2) \to 0 \quad {\mbox{{for}}} \quad Q^2 \to \infty )$
on $G_M(Q^2)$ by choosing $M=M_0$ in (16,17), or, to
put it more generally, to check the functional form of (17)
against the experimentally observed behaviour of $Q^4G(Q^2)$ for large $Q^2$
by choosing $M$ as an adjustable parameter. Inclusion of other terms in
the lagrangian $\cal L$, quantum corrections, additional degrees of
freedom, may affect the position of zeros in different FFs in different
ways. In tree approximation, for a specific choice of ${\cal L}$, we
therefore should not expect $M$ to be necessarily the same for different
formfactors.
Due to the lack of the factor $\gamma^{-2}$ on the right
hand side in (18), superconvergence cannot be imposed on $G_E$
by any choice of $M$.

The low-$Q^2$
behaviour is not strongly affected by these variations in $M$,
although due to the factor $\gamma^{-2}$ in front of $G_M^{nr}$ in (17),
even the magnetic radius receives a small contribution from finite
values of $M$.

\section{Electromagnetic form factors of the nucleon}

We have discussed in the preceding sections that the
part of the effective pionic action presently established
in the mesonic sector is not sufficient
for a realistic description of baryons in the soliton sector.
A specific set of higher-order terms is conveniently generated
through explicit inclusion of vector mesons.
The relevant coupling constants need not necessarily be in close
agreement with their values in the mesonic sector as long as we
consider observables only in tree approximation. Therefore
it may be advisable to keep
also an additional small fourth-order term of the Skyrme type (12)
in order to supply this important coupling with sufficient strength.
As a most simple effective lagrangian we therefore choose the
minimal model which comprises $\rho$ and $\omega$  mesons
together with the pionic field $U$ in chiral covariant way:
\be
{\cal L}={\cal L}^{(2)}+{\cal L}^{(4)}_A+
{\cal L}^{(\rho)}+{\cal L}^{(\omega)}
\ee
\be
{\cal L}^{(\rho)}= \int \left(-\frac{1}{8} tr \rho_{\mu\nu} \rho^{\mu\nu}
+\frac{m_\rho^2}{4} tr(\rho_\mu
-\frac{i}{2g}(l_\mu-r_\mu))^2 \right) d^3x,
\ee
\be
{\cal L}^{(\omega)}=\int \left(-\frac{1}{4} \omega_{\mu\nu}
\omega^{\mu\nu} +\frac{m_\omega^2}{2}
\omega_\mu \omega^\mu +3g_\omega \omega_\mu B^\mu \right) d^3x,
\ee
topological baryon current $B_\mu$
\be
B_\mu=\frac{1}{24 \pi^2} \epsilon_{\mu\nu\rho\sigma} tr L^\nu L^\rho
L^\sigma,
\ee
\be
l_\mu=\xi^\dagger \partial_\mu \xi, \;r_\mu=\partial_\mu \xi
\xi^\dagger \;\;\mbox{with}\;\; \xi=U^{\frac{1}{2}}.
\ee
With experimental values for $f_\pi=93$ MeV, meson masses
$m_\pi=138$ MeV, $m_\rho=770$ MeV, $m_\omega=783$ MeV,
and $g$ fixed by the KSRF relation $g=m_\rho/(2\sqrt{2} f_\pi)$ = 2.925,
${\cal L}$ still contains $g_\omega$ and  $e$ as two free parameters.
Their influence on the resulting profiles is rather similar:
a small value of $g_\omega$ can be compensated by a stronger Skyrme
term and vice versa. For all the following results
we choose $g_\omega = 4$ and $e=12$.

The contributions of the vector mesons to the electromagnetic
currents are defined through the gauge transformations (with local
gauge field $\epsilon$)
\be
\rho^\mu \rightarrow  e^{i \epsilon Q_V}
\rho^\mu e^{-i \epsilon Q_V}  +\frac{Q_V}{g} \partial^\mu \epsilon
\ee
\be
\omega^\mu \rightarrow  \omega^\mu
+\frac{Q_0}{g_0} \partial^\mu \epsilon
\ee
(with $Q_0=1/6 \ , \ Q_V=\tau_3 /2$). Within our $SU(2)$ scheme we can
allow $g_0$ to differ from $g$ and thus
exploit the freedom in the e.m. coupling of the isoscalar $\omega$-mesons.

The nucleon isoscalar FFs are of very simple form
which involves only the spatial distribution of the baryon density
$B_0(r)$ and the ratio $g_\omega/g_0$ which determines the
contribution of the $\omega$-mesons to the isoscalar part
of the e.m. current:
\be
 G^0_E (Q^2)
 =\frac{1}{2} \int d^3 r \; j_0 (Qr)\left(\frac{g_\omega}
{g_0(1+Q^2/m^2_\omega)} + 1 - \frac{g_\omega}{g_0}\right) B_0(r)
\ee
\be
 G^0_M (Q^2)
 =\frac{M_N}{2 \Theta} \int d^3 r \; \frac{j_1(Qr)}{Qr}\left(\frac{g_\omega}
{g_0(1+Q^2/m^2_\omega)} + 1 - \frac{g_\omega}{g_0}\right)r^2 B_0(r)
\ee
(The functional form of other electromagnetic FFs in this model is
given e.g. in~\cite{Mei93}.)

Comparison with the Galster parametrization~\cite{Ga71} in figs.3 and 4
\be
G^0_E(Q^2)/G_D(Q^2)=\frac{1}{2}\left( 1+\frac{0.54 Q^2}{1+1.59 Q^2}\right) \;
, \;\;\;\; G_D(Q^2)=1/(1+Q^2/0.71)^2 \;
\ee
\be
G^0_M(Q^2)/G_D(Q^2)=\mu_p+\mu_n=\mbox{const.}|_{Q^2}
\ee
shows that $g_\omega/g_0$ is not equal to one (which
would imply complete vector dominance for the isoscalar FFs).
Instead, the ratio $g_\omega/g_0$ should be around $\sim 0.55 - 0.6$.
(This holds for a sufficiently wide range of parameters $e$ and
$g_\omega$  in the lagrangian). In our case both isoscalar FFs
agree well with the Galster parametrization below $Q^2 < 1$ (GeV/c)$^2$,
both for the same value of $g_\omega /g_0 = 0.58$.
This choice then also leads to quite satisfactory agreement for
the proton electric FF $G_E^p(Q^2)$ with the data in the region
$Q^2<1$ (GeV/c)$^2$ (although the electric radius of the proton
still is slightly too big: $r_{Ep}=0.93$ fm as compared to the
observed value of 0.86 fm).

The resulting e.m. FFs for the proton (divided through the standard
dipole $G_D$) are shown in figs.5 and 6,
plotted against the logarithm of $Q^2$.
The dashed lines show the results where both, $G_E$ and $G_M/\mu_p$,
are calculated for the {\it same} value
of the kinematic mass $M=M_N=0.94$ GeV in (16).
The rapid decrease of $G_E/G_D$ above $Q^2 \sim 1$ (GeV/c)$^2$
which is in apparent conflict with the SLAC data,
has its origin in the fact that the value of $M_0$ which characterizes the
first low-lying zero of $G_E^{nr}$ according to (19) appears at
$M_0=0.72$ GeV $<M_N$. With the choice $M=M_N$ in (16) the zero in $G_E$
is pushed up to $Q^2\approx $ 5 (GeV/c)$^2$
by the boost to the Breit frame in (18).
It can be shifted to higher $Q^2$ by decreasing $M$ towards $M_0$
(the full line in fig.5 is calculated for $M=0.85$ GeV). For $M=M_0$
the zero is, of course, completely removed (i.e. shifted to infinity).
Because the rapid decrease of $G_E/G_D$ is due to a zero in $G_E^{nr}$
it cannot be removed by an additional factor
$\gamma^{-2}$ in front of $G_E^{nr}$ which may appear in bag models.

For $M=M_N$ the proton magnetic FF $G_M^p(Q^2)$ deviates significantly
from the standard dipole $G_D$ above $Q^2>1$ (GeV/c)$^2$,
(dashed line in fig.6). Because for the nonrelativistic magnetic FF
we find $M_0=1.147$ GeV $>M_N$,
the first zero in $G^{nr}_M$ is completely removed by the boost to the
Breit frame (i.e. it is shifted beyond infinity into the timelike region
of $Q^2$).
The full line in fig.6 is calculated with
$M=1.138$ GeV which is very close to the requirement of
superconvergence ($M=M_0$). Remarkably, with this small change in the
kinematic mass it is possible to obtain the impressive agreement
with the SLAC data up to 30 (GeV/c)$^2$.

For the chosen set of $g_\omega$ and $e$ the proton magnetic moment is
$\mu_p = 2.90$, and the neutron magnetic moment is $\mu_n = -2.49$.
The discrepancy of the latter with its experimental value of -1.91
is due to the fact that in soliton models the isoscalar magnetic
moment
\be
\mu_S / M_N = \frac{<r_B^2>}{3 \Theta}
\ee
is closely tied to the inverse soliton moment of inertia $\Theta$,
a relation which is {\it two} $N_c$ orders down from the ${\cal O}(N_c)$
isovector magnetic moment.
So, similar to the case of $g_A$ discussed earlier,
we expect decisive improvement only at the two-loop level.
Still it is interesting to note that the resulting Foldy term
\be
\frac{3 \mu_n}{2 M_N^2}= -0.165 \mbox{ fm}^2
\ee
essentially saturates the calculated square radius of the neutron
\be
<r_E^2>_n= -6 \; dG_E^n(0)/dQ^2 =-0.158 \mbox{ fm}^2
\ee
(the experimental value is  $<r_E^2>_n=-0.114\pm 0.003$ fm$^2$).
Apparently, the soliton model naturally reproduces the experimental
observation that the slope of the Dirac form factor of the neutron
$d F_{1n}(0)/dQ^2$ is extremely small
(experimentally it is $\sim$ 0.002 fm$^2$).

Although details will depend on the choice of parameters
in the effective lagrangian it is evident
from fig.6 that the functional form (17)
is able to describe the general pattern of the observed proton magnetic FF
over the whole range of measured $Q^2$ values if superconvergence
is imposed, without any further "QCD" corrections~\cite{QCDcorr}.
This makes it unlikely to find unambiguous signatures from additional
short-range degrees of freedom in the observed form of $G_M^p$.
A similar result was obtained by H\"ohler~\cite{Hoe76}
in terms of a suitable parametrization of the spectral function.

The electric FF is dominated by a zero in the few (GeV/c)$^2$ region
which is difficult to avoid and which appears to be in conflict with
the SLAC data. It is an interesting question whether quantum
corrections or inclusion of additional degrees of freedom
may shift the first zero in the nonrelativistic electric FF closer to
the value $(4 M_N^2)$.
The scaling relation $G_M^p(Q^2)/\mu_p =G_E^p(Q^2) $, however,
is satisfied with good accuracy up to $Q^2\approx 1$ (GeV/c)$^2$ which is
quite remarkable for a model where the Besselfunctions $j_0$ and
$j_1$ determine the electric and magnetic FFs, respectively, (which
naively implies for the ratio of the radii $<r^2_M>/<r^2_E>\sim 3/5$).
Clearly, more experimental information on the proton electric FF
in the few (GeV/c)$^2$ region will be very helpful for a critical
assessment of these generic implications of the soliton model.

\section{ The strong $\pi N N$ form factor}

Let me finally make some remarks concerning the strong form factor
which governs the $\pi N N$ interaction in soliton models.
First it should be stressed, that the analysis of
the meson-baryon scattering S-matrix in the soliton
sectors of effective meson lagrangians does not {\it require} to separately
consider meson-baryon form factors: the spatial structure of the
interaction is determined by the selfconsistently calculated soliton
profiles and automatically taken care of in the scattering
equations.  This holds, of course, also for the analysis of the
baryon-baryon interaction, or for the structure of the
deuteron or other nuclei.

It may, however, be desirable to extract meson-baryon form factors from
soliton solutions of mesonic actions, to enable a comparison
with the (freely invented) form factors typically used in
conventional meson-exchange models of the baryon-baryon interaction.
A principal difficulty lies in the fact that effective theories
allow for arbitrary unitary redefinitions of the interpolating
fields which leave the S-matrix unaffected.
It is therefore essential to extract the form factors in such a way that
they are independent of the specific definition chosen for the
interpolating field. This is indeed possible if one takes due
care of the local metric associated with a given choice of
interpolating field.

These metrical factors have been disregarded in early attempts to
relate the strong form factors to the soliton
profiles: the procedure followed in
Refs.~\cite{Coh86,KMW87} is based on the equation of
motion (EOM) for a pion field $ \vec \pi $ coupled to a
(fermionic) axial source
\be
(\Box - m_\pi^2)\pi^a(x) = J^a_5(x).
\ee
Taking matrix elements for nucleon states and using translational
invariance leads to
\be
-(q^2+m_\pi^2) <N(p')|\pi^a(0)|N(p)> = <N(p')|J^a_5(0)|N(p)>
\ee
with $q=p'-p$. The matrix element on the right-hand side defines the
form factor
$G_{\pi N N}$ through
\be
<N(p')|J^a_5(0)|N(p)> = G_{\pi N N}(q^2)\; \bar u(p') i \gamma_5 \tau^a
u(p)
\ee
while the matrixelement on the left-hand side to lowest order in
$\hbar/N_c$
is the Fourier transform of the classical meson field
\be
 <N(p')|\pi^a(0)|N(p)> = \int e^{iqx} \pi^a_{cl}(x)\, dx .
\ee

Through (36),(37), and (38) the $\pi N N$ form factor thus is expressed in
terms of the classical solution for the chiral field.
It implies that in
an EOM for the fluctuating pion field derived from any chiral
effective action (conveniently formulated in terms of a unitary
matrix field $ U=\sigma + i\, \vec \tau \cdot \vec \pi $)
 \be
(\Box - m_\pi^2)\pi^a(x) = J^a_5 [ U(x)]
\ee
the matrixelements of the functional  $J^a_5 [ U(x)] $ in baryonic
configurations may
be identified with the corresponding fermionic matrix elements of $
J^a_5(x) $.

It should be noted, however, that the EOM derived from some effective
meson action is
not immediately obtained in the form (39), because the kinetic part will
generally contain a local metric.
Only after a field redefinition to absorb this metric into the chiral
field the
correspondingly transformed source function can be compared with the
fermionic matrix
elements and the form factor. Evidently, this metric can only be
identified from the
time-derivative part of the action, because any deviation of the spatial
part from the
required structure  $\nabla^2 \pi^a$ could be absorbed into the  source
function
$ J^a_5 [ U(x)]$ without a redefinition of the field.

In terms of the Maurer-Cartan forms $L^\mu = L^\mu_a \tau_a$ (2)
the kinetic part ${\cal T }$ of the lagrangian which determines the
dynamics of the field fluctuations generally is given
by
\be
{\cal T} =- \frac{f_\pi^2}{2} \int \;L^0_a M_{ab} L^0_b \;d^3x
\ee
This also holds for effective theories which contain more than two
time derivatives in their chiral action, because ${\cal T }$ is
obtained by expanding the lagrangian to second order in the fluctuations.
In the Skyrme model and related models the classical field
configuration $ \pi^a_{cl}(x)$ which characterizes
the baryon is the hedgehog
 $U_0 = \exp (i\vec{\tau}\cdot\hat{x} F(r))$ with chiral profile
$F(r)$, rotating in isospace.
For solitons of this type the only isovector which can
appear in the metric tensor $M_{ab}$ is the pion field itself,
($\vec\pi=|\vec\pi|\hat\pi$), therefore $M_{ab}$ has to be of the form
\be
M_{ab} = M_L \hat{\pi}_a  \hat{\pi}_b + M_T (\delta_{ab}- \hat{\pi}_a
\hat{\pi}_b)
\ee
with longitudinal and transverse metrical factors $M_L$ and $M_T$
depending on $\sigma$
and $|\vec\pi|$. The metric in (40) can be removed from the
kinetic energy by redefining
\be
\tilde{L}^0_a = M^{1/2}_{ab} L^0_b.
\ee
For the hedgehog soliton $\vec{\pi}$ rotating in isospace
with angular
velocity $\vec{\Omega}$ the time derivative $\dot{\vec {\pi}}$ is
purely transverse
and we have
\be
-\tilde {L}^0_a \tilde {L}^0_a = \dot{\tilde{\pi}}_{a}
\dot{\tilde{\pi}}_{a}
\ee
with redefined field $\tilde{\pi}_{a} = \sqrt{M_T} \pi_{a}$.

Combining now eqs.(36-38) with $ \pi^a_{cl}(x)$ replaced by
$ \sqrt{M_T} \pi^a_{cl}$
the $\pi N N$ form factor in the Breit frame then is obtained as
\be
G_{\pi N N}= \frac{8\pi}{3}\frac{M_N f_\pi}{Q}
(Q^2 + m_\pi^2) \int_0^\infty \!dr \,r^2 j_1(Qr) \sqrt{M_T(r)}
\;\sin F(r)
\ee
where $M_T(r)$ derives from the effective lagrangian used to
determine the chiral profile $F(r)$.
As a typical example, we consider the standard Skyrme lagrangian.
It leads to the transverse metric to be used in (44)
\be
M_T(r)= 1+\frac{1}{e^2f_\pi^2}(F'^2+\frac{\sin^2 F}{r^2})
\ee
This is the form factor which implicitly underlies the successful
application of the soliton model to $\pi N$ elastic
scattering~\cite{WE84,WMW92,HPJ90}.
In fig.7 we compare it with conventional monopole form factors
$G_{0}(Q^2) = (\Lambda^2-m_\pi^2)/(\Lambda^2+Q^2)$
with $\Lambda$ = 0.8 and 1.7 GeV.
The contribution of the Skyrme term (with $e=4.25$) to the pionic metric
causes a qualitative change in the low-$Q^2$ behaviour of the formfactor:
the slope near $Q^2=0$ is very small and the curvature is negative.
This means that for small $Q^2$ the effective $\pi NN$ coupling strength
stays much closer to its value at $Q^2=-m_\pi^2$ than for comparable monopole
form factors. It is this feature of the Skyrme model form factor which
also improves agreement of the calculated P33 phase shifts in $\pi N$
scattering with the data in the $\Delta$-resonance
region~\cite{Ho93}. If the metric factor in (44) is omitted
(i.e. (45) replaced by $M_T(r)=1$) then the resulting FF is very close
to a soft monopole FF with $\Lambda \sim 0.8$ GeV, as originally
observed in~\cite{Coh86,KMW87}.
Soft ($\Lambda \sim 0.8$ GeV) monopole FFs fail
in the NN system, since they cut out too much of the tensor force
provided by the pion: the deuteron quadrupole moment and asymptotic
D/S state ratio and the $\epsilon_1$ mixing parameter of NN scattering
(which all depend crucially on the nuclear tensor force) come out too
small~\cite{note1}.
The very hard behaviour of the Skyrme model FF for small $Q^2$
therefore proves very helpful in standard OBE calculations.
On the other hand the very soft behaviour for
$Q^2 >$ 50 $m_\pi^2$ cuts off higher momenta much
more efficiently than typical hard monopole formfactors.
Table 1 shows some results obtained by Machleidt~\cite{Ma95} for deuteron
properties and low-energy n-p scattering with the standard monopole FF
with $\Lambda=1.7$ GeV typically used for the $\pi NN$-vertex
in the Bonn potential, and with the Skyrme model FF of eqs. (44,45), instead.
Evidently, the results are very similar.

\begin{center}
\tcaption{
Deuteron and low energy scattering parameters as predicted
by the two OBE potential models discussed in the text.
Column `Experiment' gives the empirical values.}
\begin{tabular}{llll}
\\ \hline\hline  \\
     & Bonn-B~~~~~ & Skyrme FF~~ & Experiment$^{a}$ \\
\\ \hline \\
{\bf Deuteron:} \\
Binding energy $-\epsilon_{d}$ (MeV) & 2.2246  & 2.22454 & 2.224575(9)  \\
D-state probability $P_{D}$ (\%) & 4.99 &  4.71 & -- \\
Quadrupole moment $Q_{d}$ (fm$^{2}$) & 0.278$^{b}$ & 0.274$^b$ & 0.2860(15)\\
Magnetic moment $\mu_{d}$ ($\mu_{N}$) & 0.8514$^{b}$ &      & 0.857406(1) \\
Asymptotic S-state $A_{S}$ (fm$^{-1/2}$) & 0.8860 & 0.8876 & 0.8846(8)  \\
Asymptotic D/S-state $D/S$ &   0.0264 &    0.0257 &   0.0256(4) \\
Root-mean-square radius $r_{d}$ (fm) &1.9688 &     &  1.968(5) \\
\\
\multicolumn{4}{l}{{\bf Neutron-proton low-energy scattering:}} \\
\multicolumn{4}{l}{(scattering lenght $a$,
effective range $r$)} \\
$^{1}\!S_{0}$: $a_{np}$ (fm) & --23.75 & --23.75 &  --23.748(10) \\
\hspace*{.7cm} $r_{np}$ (fm) & 2.71    & 2.73    &  2.75(5)  \\
$^{3}\!S_{1}$: $a_{t}$ (fm) &      5.424   & 5.434 &  5.419(7) \\
\hspace*{.7cm} $r_{t}=\rho (0,0)$ (fm) & 1.761 & 1.776 & 1.754(8) \\
\\ \hline\hline \\
\multicolumn{4}{l}{\footnotesize
$^{a}$ The figures in parentheses after the experimental values
give the one-standard-deviation}
\\ \multicolumn{4}{l}{\footnotesize
uncertainties in the last digits.
A comprehensive list of references for the experimental}
\\ \multicolumn{4}{l}{\footnotesize
values is given in Table 4.2
(p.~227) of Ref.~\cite{Mac89}.}
\\ \multicolumn{4}{l}{\footnotesize
$^{b}$ The meson-exchange current contributions to the moments are not
included}
\\ \multicolumn{4}{l}{\footnotesize
in the predictions.}
\end{tabular}
\end{center}

\section{Conclusion}
First attempts to relate the soliton description of baryons to the
established parts of the low-energy effective action of ChPT have shown
that for application in the soliton sector more knowledge about the
higher chiral orders in ${\cal L}$ is necessary. However, as a model with
'effective' low-energy coupling constants a truncated form of ${\cal L}$
proves very useful for studying the next-to-leading-order terms in the
$1/N_c$ expansion. It turns out that they in fact constitute sizable
contributions to physical observables and remarkably improve the
overall quality of the soliton picture. For some quantities, especially
for $g_A$, we even have to expect large next-to-next-to-leading-order
corrections. It seems not to be possible to consistently absorb these
quantum corrections into a suitable choice of effective coupling
constants for the tree approximation even with inclusion of vector
mesons.

We have presented a detailed comparison of e.m. FFs (evaluated in tree
approximation with relativistic corrections) with recent data. It
shows the capability of the soliton model to describe the peculiar
features of the FFs over a large range of $Q^2$, depending on the precise
position of the first zeros in the nonrelativistic FFs. In the model
we have used here the electric proton FF is characterized by a zero in
the few (GeV/c)$^2$ region which we find difficult to avoid. It would
be interesting to see how this feature is affected by inclusion of
further mesonic degrees of freedom, or by loop corrections. But in any case,
the quality of agreement which seems possible within these models will
make it difficult to see clear signatures from the quark structure of
nucleons in the presently available data of electro-magnetic form
factors.

\newpage
\fcaption{The scale dependence of the baryon mass $M(\mu)$ in tree +
one-loop approximation (full line). The dashed line shows the soliton
mass $M_S(\mu)$ which is implicitly scale dependent through the
renormalized coupling constants $e(\mu)$ and $\gamma(\mu)$ in the
lagrangian (12-14). (From~\cite{MW95}).}

\fcaption{The isoscalar electric FF (divided by the dipole FF
$G_D(Q^2)$) for three different values of $g_\omega/g_0 = 1$ (lowest
curve), 0.58 (middle curve), 0 (upper curve). The dashed line is the
Galster parametrization of eq.(30).}

\fcaption{The isoscalar magnetic FF (divided by the dipole FF
$G_D(Q^2)$ ) for $g_\omega/g_0 = 1$ , 0.58 , 0 (as in fig. 3). The dashed
line is the constant $\mu_0 = \mu_p + \mu_n$ . }

\fcaption{The electric FF of the proton $G^p_E/G_D$
(divided by the dipole $G_D$) for the kinematic mass $M$ = 0.94
GeV (dashed line) and $M$ = 0.85 GeV (full line). The data
points are from the compilation of refs.~\cite{Hoe76} (open circles)
and~\cite{SLAC94} (triangles). }

\fcaption{The magnetic FF of the proton $G^p_M/G_D$
(divided by the dipole $G_D$) for the kinematic mass $M$ = 0.94
GeV (dashed line) and $M$ = 1.138 GeV (full line). The data
points are from the compilation of refs.~\cite{Hoe76} (open circles),
{}~\cite{Si93} (dots), and~\cite{SLAC94} (triangles). }

\fcaption{The normalized $\pi NN$ formfactor (44) for the Skyrme model with
$e$ = 4.25 (full line). The dotted and dashed lines show the conventional
monopole form factors for $\Lambda$ = 1.7 GeV and $\Lambda$ = 0.8 GeV,
respectively.}

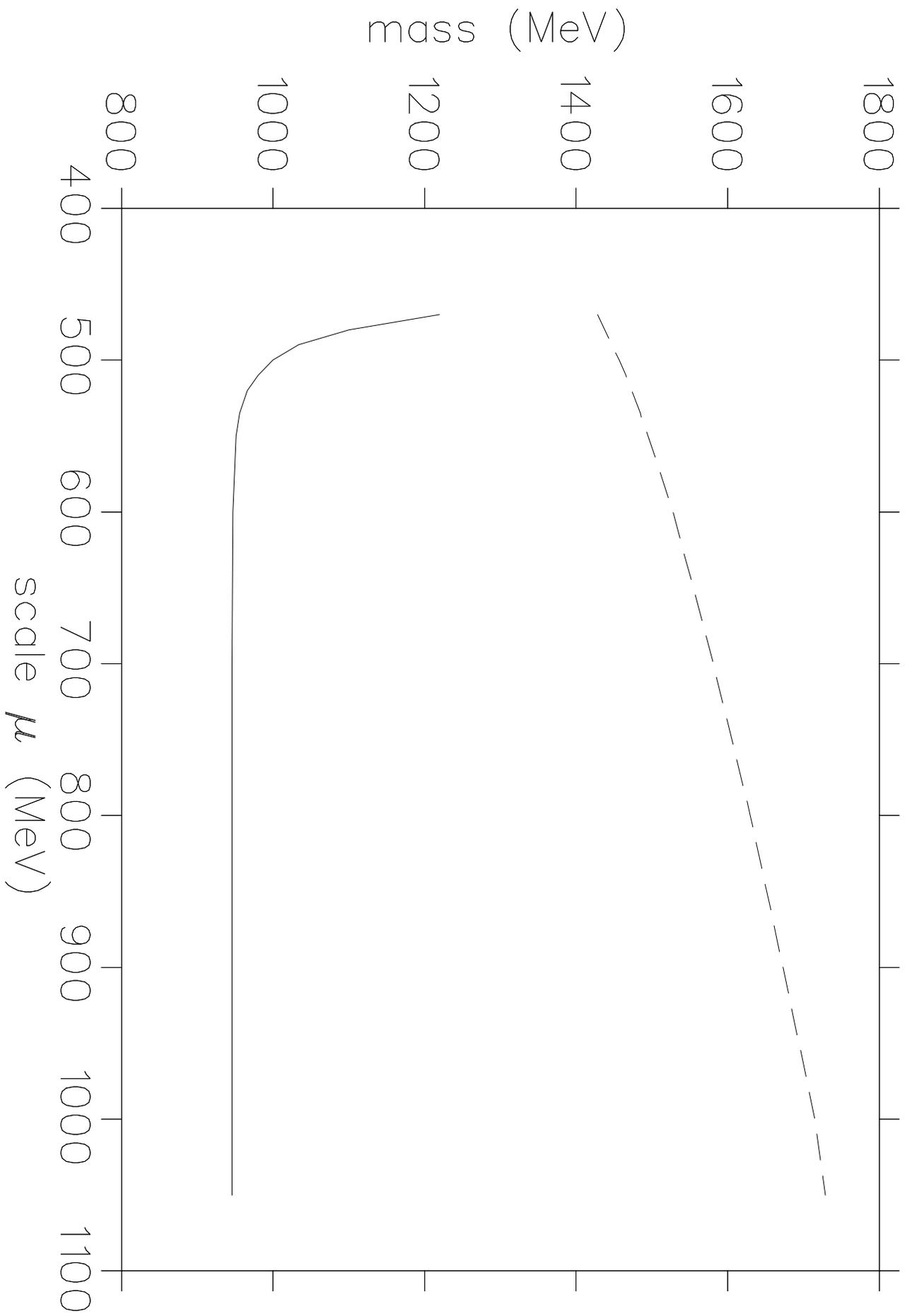
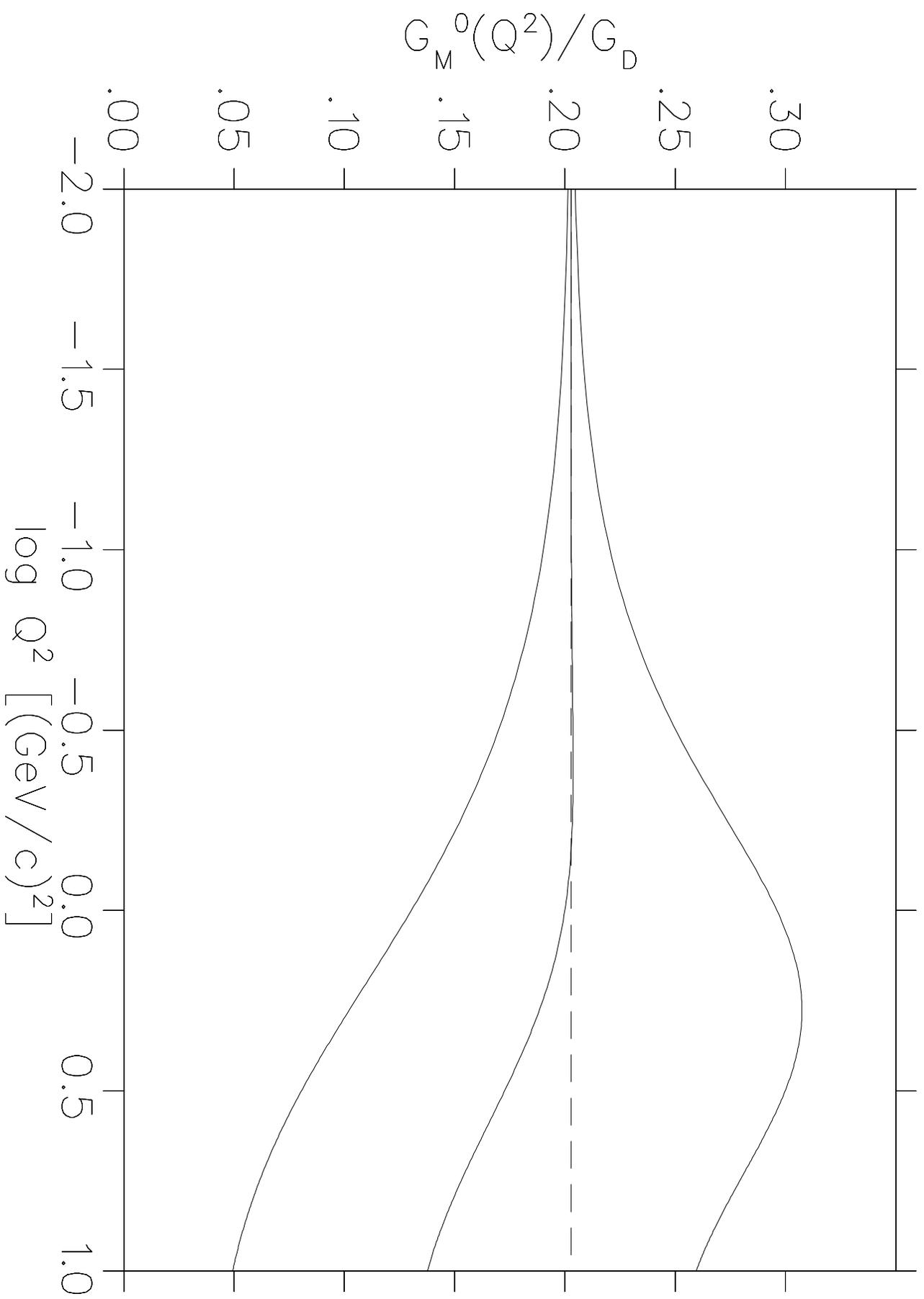
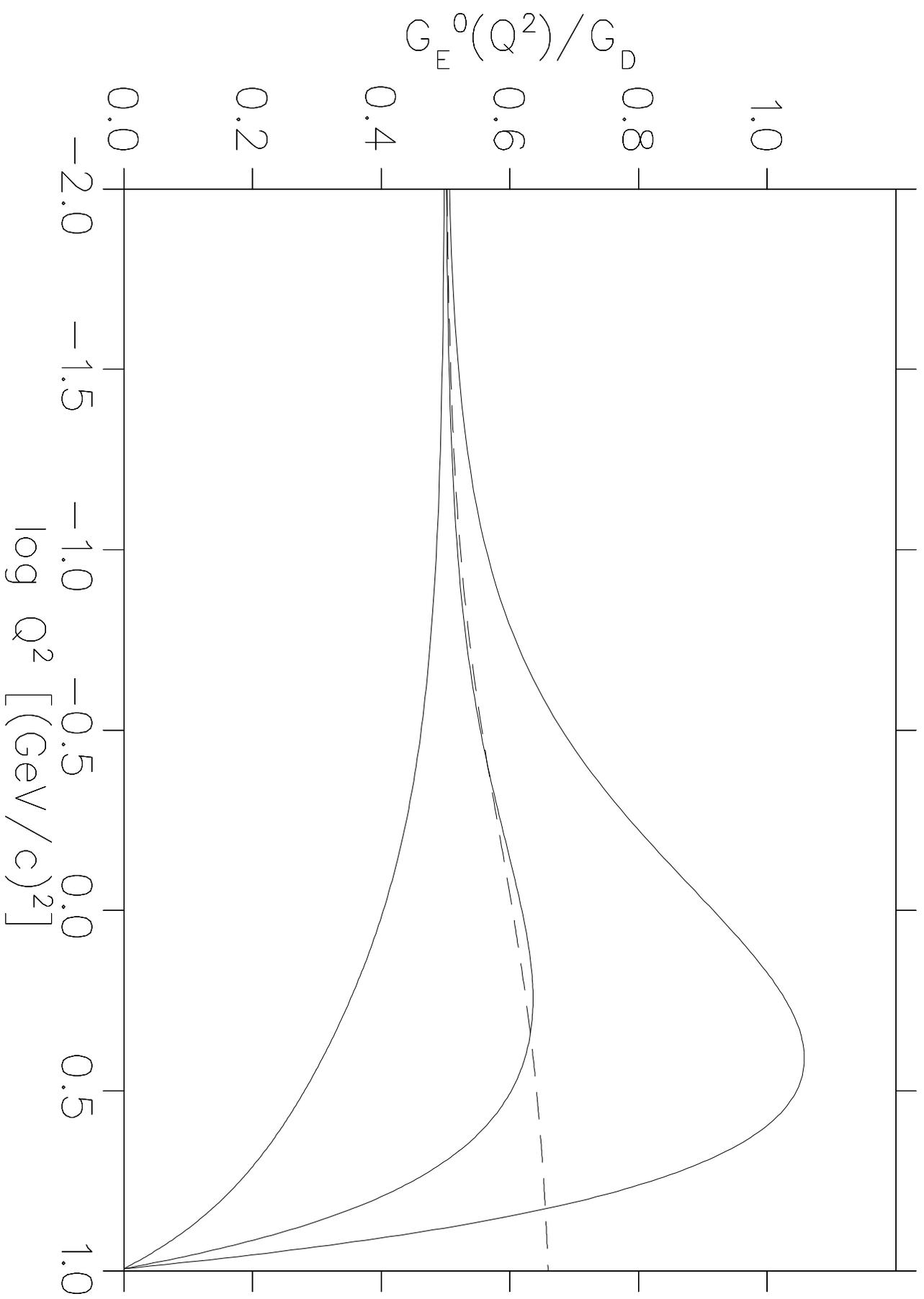
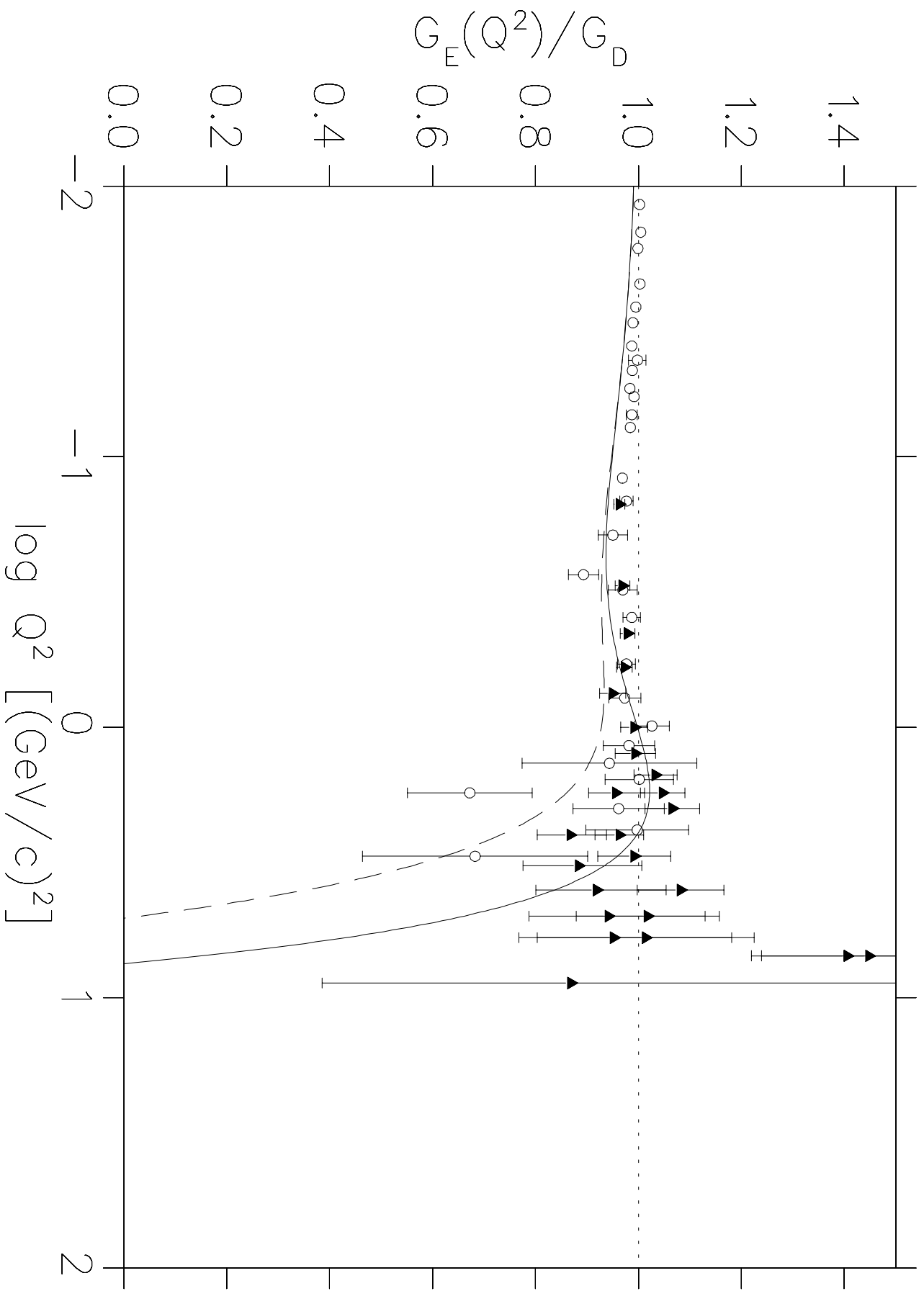
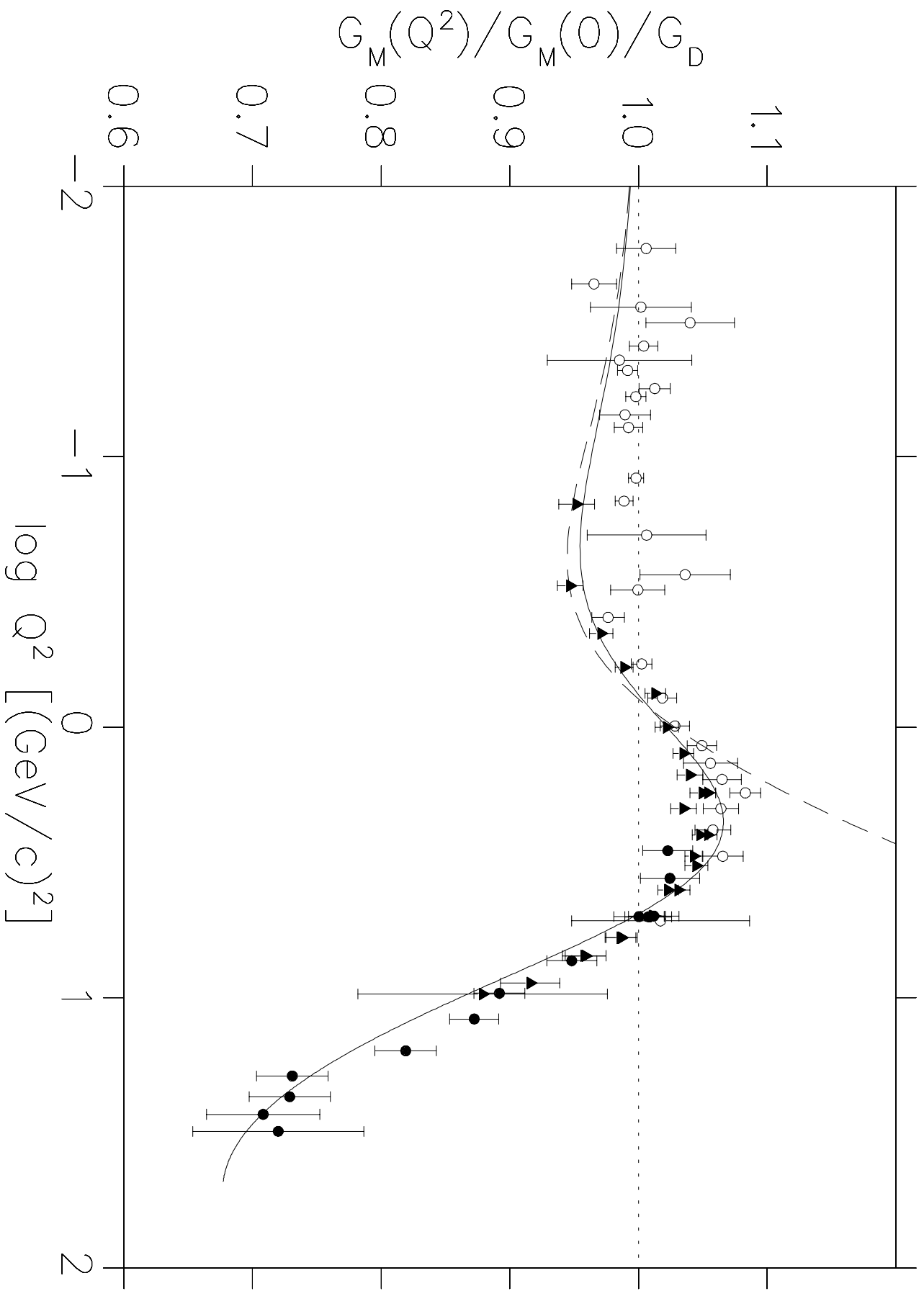
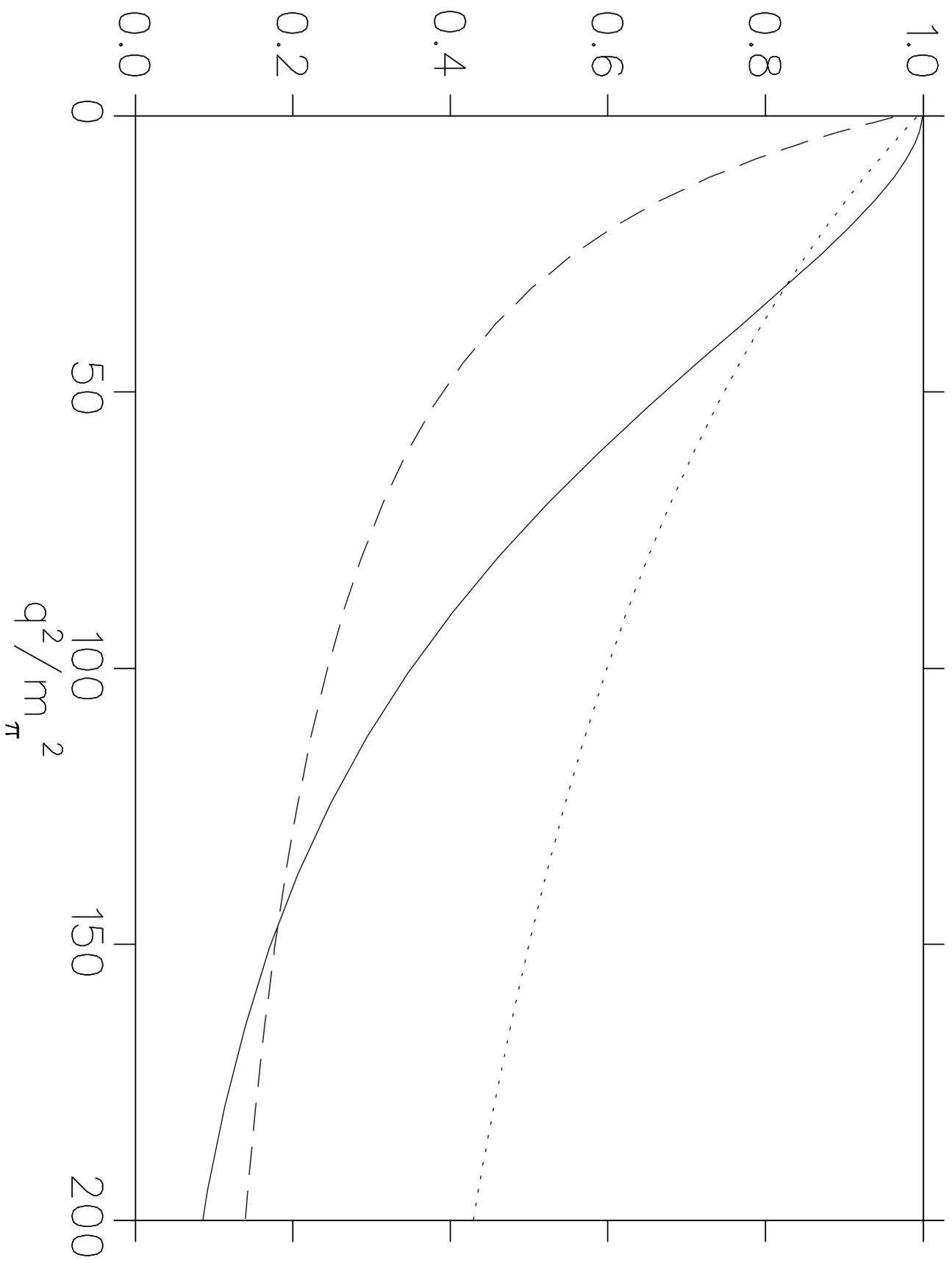
\end{document}